\documentclass[12pt]{article}
\usepackage{amssymb}
\usepackage[dvips]{epsfig}

\begin{document}

\author{R. A. C. Correa\thanks{
fis04132@gmail.com}, A. de Souza Dutra\thanks{%
dutra@feg.unesp.br}, M. B. Hott\thanks{%
marcelo.hott@pq.cnpq.br} \\
%EndAName
\\
UNESP Univ Estadual Paulista - Campus de Guaratinguet\'{a} - DFQ.\\
Av. Dr. Ariberto Pereira Cunha, 333\\
12516-410 Guaratinguet\'{a} SP Brasil}
\title{Fermion localization on degenerate and critical branes}
\maketitle

\begin{abstract}
In this work we analyze the localization of fermions on degenerate
and critical Bloch branes. This is done directly on physical
coordinates, in constrast to some works that has been using
conformal coordinates. We find the range of coupling constants of
the interaction of fermions with the scalar fields that allow us to
have normalizable fermion zero-mode localized on the brane on both,
critical and degenerate Bloch branes. In the case of critical branes
our results agree with those found in [Class. Quantum Grav.
\textbf{27} (2010) 185001]. The results on fermion localization on
degenerate Bloch branes are new. We also propose a coupling of
fermions to the scalar fields which leads to localization of
massless fermion on both sides of a double-brane.
\end{abstract}

\section{Introduction}

The idea that the Universe we live in can be realized by a static 3-domain
wall (3-brane) immersed in a (4,1)-dimensional world has opened a pathway
for the localization of matter \cite{rubakov1} and gauge bosons \cite%
{dvali-shifman} in worlds with large extra dimensions without resorting to
mechanisms of compactification \cite{akama}. Large extra dimensions has also
provided mechanisms to solve the hierarchy of interactions problem \cite%
{hamed-dvali}-\cite{randall1} as well as the cosmological constant problem
\cite{rubakov2}. In \cite{randall2} it was shown that the effective
gravitational potential between two particles recovers the Newtonian
behavior, since one has localization of gravitons on a thin brane in
five-dimensional space-time with an warped geometry and the cosmological
constant is related to the brane tension. Later the localization of matter
(spin-zero, spin-1/2 and spin-3/2) in the Randall-Sundrun framework was
shown to be possible, under certain conditions over the brane tension \cite%
{gabadadze}. It is important to emphasize that the localization of fermions
on thin branes is, in fact, provided by an \textit{ad hoc} soliton and the
mechanism for localization follows straightforwardly \textit{\`{a} la}
Jackiw and Rebbi \cite{jackiw-rebbi}. As a matter of fact, it is such a
soliton that \ provides the domain-wall and the fermion localization in the
scenario proposed by Rubakov and Shaposhnikov \cite{rubakov1}.

By introducing a non-linear model with a set of scalar fields, in
5-dimensional space-time with warped\ geometry, one has a set of coupled
non-linear differential equations whose minimum energy solutions are thick
branes and self-consistent warp factors. The thick branes also separate the
space in two patches characterized by a peculiar warp factor whose
asymptotic behavior is an anti-de Sitter (AdS$_{5}$) space \cite{wise}-\cite%
{giovannini}, as in the Randall-Sundrun framework (thin brane), but without
singularities. The stability of such solutions is very difficult to be
proven due to the intricate differential equations one has to deal with \cite%
{dewolfe2}. Notwithstanding, one still has localization of gravitons on
thick branes, as has been shown in \cite{gremm}. In fact, this program has
been developed as a generalization of the domain wall universe, by taking
into account the stabilization of gravity fluctuations via domain-walls in
supergravity theories \cite{cvetic1}.

Localization of matter on thick branes has been illustrated by using several
different non-linear models for scalar fields coupled with gravity \cite%
{koley}-\cite{slatyervolkas}. One of those models is the Bloch branes model
\cite{bazeia-gomes}, which comprises two interacting scalar fields whose
classical solutions are the Bogomolnyi-Prasad-Sommerfield (BPS) and the warp
factor can also be obtained as solution of a first-order differential
equation. Moreover, the model exhibits a richer structure due to the variety
of kinks (solitons) it comprises \cite{shifman}-\cite{dutraplb}, leading to
what has been called degenerate and critical Bloch branes \cite%
{dutrahottamaro}, with a self-consistent warp factor and localization of
gravitons. A natural track that has been followed is the localization of
matter in such a variety of branes.

In \cite{almeida} the localization of massless fermions has been studied
together with an analysis of resonant fermion modes with a specific coupling
of fermions to the classical BPS configurations of Bloch branes and in \cite%
{china} it is studied the localization of massless fermions on critical
Bloch branes. It has been explicitly pointed out \cite{dutrahottamaro} that
critical Bloch branes arise as a critical limit of degenerate Bloch branes
due to a running, but limited, constant of the integration of the orbit
equation relating the classical configurations for the scalar fields. Thus
the localization of massless fermions on degenerate and critical branes is
an important question to be analyzed. As a matter of fact, a Bloch brane may
be seen as a thick brane that evolutes to a thicker one, namely, a
degenerate Bloch brane which, in its turn, splits into two branes whose
separation becomes large as the degeneracy parameter approaches the critical
value, at which the critical brane is triggered.  Such a picture resembles
the description of first order phase transitions which was also used in the
context of brane worlds to describe brane splitting \cite{campos}. In fact,
as a counterpoint, in the model we use here, the free energy itself does not
depend on the temperature; instead, this dependence is implicit in the
degeneracy parameter.

The objective of the present work is to analyze the fate of massless
fermions trapped on a split brane. In section II we give a brief review of
the model to be used, together with the consistent branes and warp factors
solutions. In section III we deal with the localization of massless fermions
on the critical and degenerate branes and the IV section is devoted to the
conclusions and further remarks.

\section{A brief review of a model and its degenerate and critical Bloch
branes}

The action in five-dimensional gravity coupled to two interacting real
scalar fields, can be represented by

\begin{equation}
S=\int d^{4}xdr\sqrt{\left\vert g\right\vert }\left[ -\frac{1}{4}R+\frac{1}{2%
}(\partial _{\mu }\phi \partial ^{\mu }\phi +\partial _{\mu }\chi \partial
^{\mu }\chi )-V(\phi ,\chi )\right] ,  \label{1.1}
\end{equation}

\noindent where $g\equiv \det (g_{ab})$ and

\begin{equation}
ds^{2}=g_{ab}dx^{a}dx^{b}=e^{2A(r)}\eta _{\mu \nu }dx^{\mu }dx^{\nu
}-dr^{2},\ \ (a,b=0,....,4),  \label{1.2}
\end{equation}

\noindent where $r$ is the extra dimension, $\eta _{\mu \nu }$ the usual
Minkowski metric in the four space-time dimensions and $e^{2A(r)}$ is the
so-called warp factor.

If the potential $V(\phi ,\chi )$ can be written in terms of a
superpotential as

\begin{equation}
V(\phi ,\chi )=\frac{1}{2}\left[ \left( \frac{\partial W(\phi ,\chi )}{%
\partial \phi }\right) ^{2}+\left( \frac{\partial W(\phi ,\chi )}{\partial
\chi }\right) ^{2}\right] -\frac{4}{3}W(\phi ,\chi )^{2},  \label{1.7}
\end{equation}%
By substituting the superpotential, and using the orbit equation, we obtain

\begin{eqnarray}
A(\chi ) &=&\alpha _{0}+\left( \frac{2\lambda a^{2}}{9\mu }\right) \ln (\chi
)-\frac{1}{9}\left( \frac{\lambda -3\mu }{\lambda -2\mu }\right) \chi
^{2}-\left( \frac{c_{0}}{9}\right) \chi ^{\lambda /\mu },{\hspace{0.7cm}}%
(\lambda \neq 2\mu ),  \label{2.8} \\
&&  \nonumber \\
A(\chi ) &=&\alpha _{1}+\left( \frac{2\lambda a^{2}}{9\mu }\right) \ln (\chi
)-\frac{\left( 3\mu +\lambda c_{1}\right) }{18\mu }\chi ^{2}-\frac{1}{6\mu }%
\chi ^{2}\left( \ln (\chi )-\frac{1}{2}\right) {\hspace{0.7cm}}(\lambda
=2\mu ),  \label{2.9}
\end{eqnarray}

\noindent where $\alpha _{0}$ and $\alpha _{1}$ are arbitrary integration
constants, which are chosen to be $A(r=0)=0$.

It has also been found in  \cite{dutraplb, dutrahottamaro} that the
classical solutions for $c_{0}<-2a$ and $\lambda =\mu $ are given by

\begin{eqnarray}
\chi _{DBW}^{(1)}(r) &=&\frac{2a^{2}}{\left( \sqrt{c_{0}^{2}-4a^{2}}\right)
\cosh (2\mu ar)-c_{0}},  \label{2.88} \\
&&  \nonumber \\
\phi _{DBW}^{(1)}(r) &=&\frac{a\left( \sqrt{c_{0}^{2}-4a^{2}}\right) \sinh
(2\mu ar)}{\left( \sqrt{c_{0}^{2}-4a^{2}}\right) \cosh (2\mu ar)-c_{0}}.
\label{2.89}
\end{eqnarray}

The corresponding warp factor is expressed as

\begin{eqnarray}
e^{2A(r)} &=&N\left[ \frac{2a^{2}}{\left( \sqrt{c_{0}^{2}-4a^{2}}\right)
\cosh (2\mu ar)-c_{0}}\right] ^{4a^{2}/9}\times  \nonumber \\
&&  \nonumber \\
&&\times \exp \left\{ \frac{2a^{2}\left[ c_{0}^{2}\pm 4a^{2}-c_{0}\left(
\sqrt{c_{0}^{2}-4a^{2}}\right) \cosh (2a\mu r)\right] }{9\left[ \left( \sqrt{%
c_{0}^{2}-4a^{2}}\right) \cosh (2a\mu r)-c_{0}\right] ^{2}}\right\} ,
\label{2.90}
\end{eqnarray}

\noindent where $N$ is chosen such that $e^{2A(0)}=1$, for plotting
convenience.

On the other hand, for $\lambda =4\mu $ and $c_{0}<1/(16a^{2})$ the
solutions can be written as

\begin{eqnarray}
\chi _{DBW}^{(2)}(r) &=&-\frac{2a}{\sqrt{\left( \sqrt{1-16c_{0}a^{2}}\right)
\cosh (4\mu ar)+1}},  \label{2.91} \\
&&  \nonumber \\
\phi _{DBW}^{(2)}(r) &=&\frac{a\left( \sqrt{1-16c_{0}a^{2}}\right) \sinh
(4\mu ar)}{\left( \sqrt{1-16c_{0}a^{2}}\right) \cosh (4\mu ar)+1}.
\label{2.92}
\end{eqnarray}

\noindent with the warp factor

\begin{eqnarray}
e^{2A(r)} &=&N\left[ -\frac{2a}{\sqrt{\left( \sqrt{1-16c_{0}a^{2}}\right)
\cosh (4\mu ar)+1}}\right] ^{16a^{2}/9}\times  \nonumber \\
&&  \nonumber \\
&&\times \exp \left\{ -\frac{4a^{2}}{9}\left[ \frac{1+8a^{2}c_{0}+\left(
\sqrt{1-16a^{2}c_{0}}\right) \cosh [4\mu ar]}{\left( 1+\left( \sqrt{%
1-16a^{2}c_{0}}\right) \cosh [4\mu ar]\right) ^{2}}\right] \right\} .
\label{2.93}
\end{eqnarray}

The set of solutions above, was baptized by Dutra and Hott \cite%
{dutrahottamaro} as degenerate Bloch walls (DBW).

Furthermore, an interesting class of analytical solutions, named as critical
Bloch walls (CBW), was shown to exist when the constant of integration
equals a critical value. For $\lambda =\mu $ and $c_{0}=-2a$, one has the
set of solutions for the scalar fields

\begin{eqnarray}
\chi _{CBW}^{(1)}(r) &=&\frac{a}{2}\left[ 1\pm \tanh (\mu ar)\right] ,
\label{2.94} \\
&&  \nonumber \\
\phi _{CBW}^{(1)}(r) &=&-\frac{a}{2}\left[ \tanh [\mu ar)\mp 1\right] ,
\label{2.95}
\end{eqnarray}

\noindent which leads to the following warp factor

\begin{eqnarray}
e^{2A(r)} &=&N\left\{ \frac{a}{2}\left[ 1\pm \tanh (\mu ar)\right] \right\}
^{2a^{2}/9}\times  \nonumber \\
&&  \nonumber \\
&&\exp (\frac{a^{2}}{9}[1-\tanh ^{2}(\mu ar]).  \label{2.96}
\end{eqnarray}%
For $\lambda =4\mu ~$and $c_{0}=1/(16a^{2})$ the solutions for the fields
are given by

\begin{eqnarray}
\chi _{CBW}^{(2)}(r) &=&\sqrt{2}{\ }a\frac{\cosh (\mu ar)\pm \sinh (\mu ar)}{%
\sqrt{\cosh (2\mu ar)}},  \label{2.97} \\
&&  \nonumber \\
\phi _{CBW}^{(2)}(r) &=&\frac{a}{2}\left[ \pm 1-\tanh (2a\mu r)\right] ,
\label{2.98}
\end{eqnarray}%
and the warp factor is

\begin{eqnarray}
e^{2A(r)} &=&N\left[ {\ }\frac{2~a^{2}~\mathrm{e}^{\pm 2\mu ar}}{\cosh (2\mu
ar)}\right] ^{8a^{2}/9}\times  \nonumber \\
&&  \nonumber \\
&&\times \exp \left\{ \frac{2a^{2}}{9}\frac{~\mathrm{e}^{\pm 2\mu ar}}{\cosh
(2\mu ar)}\left[ 1+\frac{~\mathrm{e}^{\pm 2\mu ar}}{4\cosh (2\mu ar)}\right]
\right\} .  \label{2.99}
\end{eqnarray}

In Fig.1 are shown profiles of the warp factor in the case of DBW with $%
\lambda =4\mu ~$for some values of the constant of integration $c_{0}$ and
in Fig. 2 the behavior of the warp fact in the case CBW and $\lambda =4\mu $
is shown for some values of the parameter $a$. \

We would like to warn the reader that the expressions for the warp factor
and the Figures 2 and 3 presented in \cite{dutrahottamaro} do not agree with
those presented here. The correct expressions and Figures are those
presented here. Despite this mistake, the conclusions of the work \cite%
{dutrahottamaro} are not wrong, except for the fact that we had attributed
the appearance of two peaks in the warp factor, presented in figure 2 of
that work, as a sign of the formation of two domain walls. As a matter of
fact, two-kink solutions can be seen from the expressions (\ref{2.89}) and (%
\ref{2.92}) for values of $c_{0}$ close to the critical value. In Fig 3. we
illustrate the two-kink solution with $\lambda =4\mu $ for two values of the
constant of integration $c_{0}$. In brane cosmology scenario we are
interested in, the formation of two branes leads to an almost flat
space-time region between them, that is, the warp factor does vary
appreciably between the two branes, as can be depicted from Figure 1. The
critical brane could be seen as two branes, but infinitely separated from
each other, such that one has a complete wetting.

The two-kink solutions of this model were recently investigated to discuss
the phenomenon of brane splitting by means of an effective model with only
one scalar field \cite{chumbeshott}. In fact, that effective model was built
based on the very same model explored here . The constant of integration $%
c_{0}$ plays the role of a coupling constant in the effective potential
obtained in \cite{chumbeshott}, such that one could think of the effective
potential as a free energy that describes a first-order phase transition
characterized by the emergence of a growing wet (disordered) phase in
between two ordered phases \cite{campos}. The branes are the domain walls
separating the disordered domain from the ordered ones, as can be seen from
the behavior of the energy density of the matter fields.

\section{Localization of fermions}

In this section, we study the localization of massive fermions on the
degenerate and critical Bloch branes \cite{dutrahottamaro}. For this, we
consider a Dirac spinor field coupled with the scalar fields by a general
Yukawa coupling. Thus, the action we are going to work is given by

\begin{equation}
S_{1/2}=\int d^{5}x\sqrt{-g}\left( \bar{\Psi}i\Gamma ^{a}D_{a}\Psi -\eta
\bar{\Psi}F(\phi ,\chi )\Psi \right) ,  \label{1}
\end{equation}%
consequently the equation of motion is

\begin{equation}
\lbrack i\Gamma ^{a}D_{a}-\eta F(\phi ,\chi )]\Psi =0,  \label{3}
\end{equation}%
where $F(\phi ,\chi )$ is a functional of the classical configurations which
are solutions of the equations (\ref{1.6}). The equation of motion for the
fermion can be rewritten as

\begin{equation}
\lbrack i\Gamma ^{\mu }D_{\mu }+i\Gamma ^{4}D_{4}-\eta F(\phi ,\chi )]\Psi
=0.  \label{4}
\end{equation}%
The relations between the warped-space gamma matrices ($\{\Gamma ^{a},\Gamma
^{b}\}=2g^{ab}$), with $g^{ab}$defined in (\ref{1.2}), and the Minkowskian
ones ($\{\gamma ^{\mu },\gamma ^{\nu }\}=2\eta ^{\mu \nu }$) can be realized
as follows

\begin{equation}
\Gamma ^{\mu }=e^{-A(r)}\gamma^{\mu }{\ } \textrm{and} {\ }\Gamma
^{4}=-i\gamma ^{5}.  \label{5}
\end{equation}%
Moreover, we have the following expression for the covariant derivative

\begin{equation}
D_{a}=(\partial _{a}+\omega _{a})=\partial _{a}+\frac{1}{4}\omega _{a}^{\bar{%
a}{\ }\bar{b}}\Gamma _{\bar{a}}\Gamma _{\bar{b}},  \label{6}
\end{equation}

\noindent where $\bar{a}$ and $\bar{b}$, denote the local Lorentz indices.
Thus the spin connection $\omega _{a}^{\bar{a}{\ }\bar{b}}$ is given by

\begin{equation}
\omega _{a}^{\bar{a}{\ }\bar{b}}=\frac{1}{2}E^{b{\ }\bar{a}}(\partial
_{a}E_{b}^{\bar{b}}-\partial _{b}E_{a}^{\bar{b}})-\frac{1}{2}E^{b{\ }\bar{b}%
}(\partial _{a}E_{b}^{\bar{a}}-\partial _{b}E_{a}^{\bar{a}})-\frac{1}{2}E^{c{%
\ }\bar{a}}E^{p{\ }\bar{b}}(\partial _{c}E_{p{\ }\bar{q}}-\partial _{p}E_{c{%
\ }\bar{q}})E_{a}^{\bar{q}}.  \label{7}
\end{equation}

\noindent In the above definition $E_{a}^{\bar{a}}$ is the vielbein, and the
non-vanishing components of $\omega _{a}$ are

\begin{equation}
\omega _{\mu }=\frac{1}{2}e^{A}(\partial _{r}A)\gamma _{\mu }\gamma _{5}.
\label{8}
\end{equation}

Then the equation of motion for the fermion is given by

\begin{equation}
\{i\gamma ^{\mu }\partial _{\mu }+e^{A(r)}\gamma _{5}[\partial
_{r}+2\partial _{r}A(r)]-\eta e^{A(r)}F(\phi ,\chi )\}\Psi (x,r)=0.
\label{9}
\end{equation}

From now on we make use the general chiral decomposition

\begin{equation}
\Psi (x,r)=\sum_{n}\psi _{Ln}(x)\alpha _{Ln}(r)+\sum_{n}\psi _{Rn}(x)\alpha
_{Rn}(r),  \label{10}
\end{equation}

\noindent with $\gamma ^{5}\psi _{Ln}(x)=-\psi _{Ln}(x)$ and $\gamma
^{5}\psi _{Rn}(x)=\psi _{Rn}(x)$. Furthermore, we assume that $\psi _{Ln}(x)$
and $\psi _{Rn}(x)$ satisfy the 4-dimensional massive Dirac equations

\begin{eqnarray}
i\gamma ^{\mu }\partial _{\mu }\psi _{Ln}(x) &=&m_{n}\psi _{Rn}(x),
\label{11} \\
&&  \nonumber \\
i\gamma ^{\mu }\partial _{\mu }\psi _{Rn}(x) &=&m_{n}\psi _{Ln}(x).
\label{12}
\end{eqnarray}

Thus, applying the chiral decomposition (\ref{10}) in equation (\ref{9}) and
using equations (\ref{11}) and (\ref{12}), we arrive at the equations below

\begin{eqnarray}
\left[ \frac{d}{dr}+2\frac{dA(r)}{dr}-\eta F(\phi ,\chi )\right] \alpha
_{Rn}(r) &=&-m_{n}e^{-A(r)}\alpha _{Ln}(r),  \label{13} \\
&&  \nonumber \\
\left[ \frac{d}{dr}+2\frac{dA(r)}{dr}+\eta F(\phi ,\chi )\right] \alpha
_{Ln}(r) &=&m_{n}e^{-A(r)}\alpha _{Rn}(r).  \label{14}
\end{eqnarray}

\noindent for the $r$-dependent parts of the spinor $\Psi (x,r)$. In order
to find reliable results concerning the fermion localization in the brane we
make use of the following ortonormalization relations for the $r$-dependent
parts of the spinor:

\begin{eqnarray}
\int_{-\infty }^{\infty }e^{3A(r)}\alpha _{Ln}(r)\alpha _{Lm}(r)dr
&=&\int_{-\infty }^{\infty }e^{3A(r)}\alpha _{Rn}(r)\alpha _{Rm}(r)dr=\delta
_{nm},  \label{14.1} \\
&&  \nonumber \\
\int_{-\infty }^{\infty }e^{3A(r)}\alpha _{Ln}(r)\alpha _{Rm}(r)dr &=&0.
\label{14.2}
\end{eqnarray}

By redefining the $r$-dependent parts of the spinor as

\begin{equation}
\alpha _{Rn}(r)=e^{-2A(r)}L_{Rn}(r),~~\alpha _{Ln}(r)=e^{-2A(r)}L_{Ln}(r),
\label{16}
\end{equation}%
we are able to get rid of the second-term in the left-hand sides of the
equations (\ref{13}) and (\ref{14}). Thus we obtain

\begin{eqnarray}
\left[ \frac{d}{dr}-\eta F(\phi ,\chi )\right] L_{Rn}(r)
&=&-m_{n}e^{-A(r)}L_{Ln}(r),  \label{19} \\
&&  \nonumber \\
\left[ \frac{d}{dr}+\eta F(\phi ,\chi )\right] L_{Ln}(r)
&=&m_{n}e^{-A(r)}L_{Rn}(r).  \label{20}
\end{eqnarray}

We notice that the above equations are equivalent to the equations for the
components of a spinor describing a massless fermion in 1+1 dimensions
subject to a mixing of scalar and vector potentials. The time-independent
equation for a fermion under such potentials can be written as $H\psi
(r)=E\psi (r)$, with the Dirac Hamiltonian given (in natural units) by $%
H=\sigma _{2}p+V_{s}(r)\sigma _{1}+V_{v}(r)$, where $p=-id/dr$ is the
momentum operator, $\sigma _{1}$ and $\sigma _{2}$ are the two non-diagonal
Pauli matrices, $V_{s}(r)=$ $-\eta F(\phi ,\chi )$ (note that $F(\phi ,\chi
) $ is a function of $r$) is the scalar potential and $%
V_{v}(r)=m_{n}e^{-A(r)}$ is the vector potential. In this analogy, one can
say that $L_{Ln}(r)$ and $L_{Rn}(r)$ play the role of the upper and lower
components for the fermion zero-mode in 1+1 dimensions. We have mentioned
that the fermion in 1+1 dimensions is massless, but in fact, the scalar
potential can be thought as a position-dependent mass. As one knows, many
examples of such systems were already solved in the literature \cite{castro}%
, particularly when the scalar potential is proportional to the vector
potential, namely $V_{s}(r)=\delta V_{v}(r)$, that allow for fermion bound
states. The vector potential can be attractive for fermions whilst it is
repulsive for antifermions and \textit{vice-versa}, that is, one can have
pair-production. On the other hand, if $\delta \geq 1$ and the vector
potential is attractive for anti-fermions, the mixing of such potentials
supports bound states. This could be explained due to an increasing of the
threshold for the pair-production provided by $V_{s}(r)$, since it
contributes as a variable mass for the fermion and the energy provided by
the electric field does not reach two times the effective mass of the
fermion. In the brane world scenario we are considering here, massive ($%
m_{n}\neq 0$) as well as massless ($m_{n}=0$) fermions might be localized
inside the brane, depending on the shape and strength of the coupling with
the scalar field $\eta F(\phi ,\chi )$, although, the issue of localization
of massive modes in the brane is a very difficult as discussed below.

Concerning the localization of massless fermions in the brane, one finds
that the $r$-dependent coefficients appearing in the chiral decomposition (%
\ref{14}) are written as

\begin{eqnarray}
\alpha _{R~_{0}}(r) &=&N_{R~_{0}}\exp [-2A(r)+\eta \int^{r}F(r^{\prime
})dr^{\prime }]\;  \nonumber \\
\alpha _{L_{\ 0}}(r) &=&N_{L~_{0}}\exp [-2A(r)-\eta \int^{r}F(r^{\prime
})dr^{\prime }]~,  \label{20.a1}
\end{eqnarray}%
where we have defined $F(r)=F(\phi (r),\chi (r))$. In general, one resorts
to an analytical function $F(\phi (r),\chi (r))~$\ in order to have
localized massless fermions in the brane and those localized states have
invariably a well defined chirality, which depends on the behavior of $%
\int^{r\rightarrow \pm \infty }F(r^{\prime })dr$ and on the sign of $\eta $,
since the normalization condition for the $r$-dependent parts of the
massless fermion are given by
\begin{equation}
\left\vert N_{R~_{0}}\right\vert ^{2}\int_{-\infty }^{\infty }e^{-A(r)+2\eta
\int^{r}F(r^{\prime })dr^{\prime }}dr=\left\vert N_{L~_{0}}\right\vert
^{2}\int_{-\infty }^{\infty }e^{-A(r)-2\eta \int^{r}F(r^{\prime })dr^{\prime
}}dr=1.  \label{20.b}
\end{equation}%
An interesting approach of this, can be found in the work by Slatyer and
Volkas \cite{slatyervolkas}. In general, if $\alpha _{R~_{0}}(r)$ is
normalizable, $\alpha _{L_{\ 0}}(r)$ is not and \textit{vice-versa, }then
the non-normalizable contribution equals to zero. This could explain why we
observe neutrinos with only one chirality in our universe. The fact that
only one of the chiralities is normalizable can be seen from the effective
Schr\"{o}dinger equation that can be derived for $L_{R0}(r)$ and $L_{L0}(r)$
which are zero-modes of effective potentials that are supersymmetric
partners of each other, and the normalization of both components would imply
into the supersymmetry breaking. One can also check that the factor $%
e^{-A(r)}$ in the integrands above is also important for the normalization
of the spinor, although it is not decisive for the resolution of the
chirality, since it amounts to the same weight in the normalization for both
chiralities.

For $m_{n}\neq 0$ we observe that the left and right components can be
decoupled. It can be shown that $L_{Ln}(r)$ obeys the following second-order
differential equation

\begin{equation}
L_{Ln}^{\prime \prime }+A^{\prime }L_{Ln}^{\prime }+[\eta A^{\prime }F+\eta
F^{\prime }-\eta ^{2}F^{2}+m_{n}^{2}e^{-2A}]L_{Ln}=0,  \label{21}
\end{equation}%
where the prime denotes derivative with respect to $r$. With the redefinition

\begin{equation}
L_{Ln}(r)=e^{-A(r)/2}f_{Ln}(r).  \label{22}
\end{equation}%
one finds that $f_{Ln}(r)~$obeys a time-independent Schr\"{o}dinger equation

\begin{equation}
-f_{Ln}^{\prime \prime }+U_{eff}^{L}~f_{L}(r)=0,  \label{23}
\end{equation}
with eigenvalue equals zero and the effective potential%
\begin{equation}
U_{eff}^{L}(r)=\eta ^{2}F^{2}-\eta F^{\prime }-\eta A^{\prime
}F+(1/4)A^{\prime 2}+(1/2)A^{\prime \prime }-m_{n}^{2}e^{-2A(r)}.  \label{24}
\end{equation}

\noindent For the right-component one finds

\begin{equation}
-f_{Rn}^{\prime \prime }+U_{eff}^{R}~f_{Rn}(r)=0,  \label{25}
\end{equation}

\noindent with

\begin{equation}
U_{eff}^{R}(r)=\eta ^{2}F^{2}+\eta F^{\prime }+\eta A^{\prime
}F+(1/4)A^{\prime 2}+(1/2)A^{\prime \prime }-m_{n}^{2}e^{-2A(r)}.  \label{26}
\end{equation}

We have reduced the problem of localization of massive fermions on a brane
into a Sturm-Liouville problem, which agrees with the results found in \cite%
{koley} and \cite{slatyervolkas}, without resorting to the transformation of
variable as done in others papers.

One can note that there is a symmetry relating the effective potentials,
namely $U_{eff}^{R}=\left. U_{eff}^{L}\right\vert _{\eta \rightarrow -\eta }$
and that the eigenstates $f_{Ln}(r)$ and $f_{Rn}(r)$ are subject to distinct
effective potentials. In order to have localization of a specific massive
mode $m_{n}$, each one of the chiralities must be a bound state of its
corresponding potential with energy equal to zero. Due to the above symmetry
and the dependence on the mass in the expressions (\ref{24}) and (\ref{26})
one has a specific potential for each massive mode and specific chirality.
In general, the equations (\ref{23}) and (\ref{25}) are very difficult to be
analytically solved, but there are two particular and unphysical cases,
namely $\eta F=m_{n}e^{-A(r)}$ and $\eta F=-m_{n}e^{-A(r)}$, for which one
can find exact expressions for $\alpha _{Rn}(r)$ and $\alpha _{Ln}(r)$,
namely $\alpha _{Rn}(r)=\alpha _{Ln}(r)\sim \exp [-2A(r)]$, which are not
normalizable.

Now, in order to compare our results with a previous one\ \cite{china}, we
address the issue of possible localization of massless fermions on the brane
by setting the general coupling

\begin{equation}
F(\phi ,\chi )=\omega _{1}\phi +\omega _{2}~\chi +\omega _{3}~\phi \chi .
\label{35}
\end{equation}%
to the fermions, where the $\omega _{i}$'s are constant parameters to be
determined such that $\alpha _{R~_{0}}(r)$ or $\alpha _{L_{\ 0}}(r)$ is
normalizable. We have analyzed the scenarios of both degenerate and critical
branes for $\lambda =\mu $.

It is important to remark that the factor $e^{-A(r)}~$(see eq. (\ref{20.b}))
diverges for $r\rightarrow \pm \infty $ in the case of DBW, such that it
does impose restrictive conditions on the normalization of the wave
functions. On the other hand, in the case of CBW, the factor $e^{-A(r)}~$
diverges for $r\rightarrow -\infty $ and is constant for $r\rightarrow
\infty $ (we have taken the upper signs in expressions (\ref{2.94})-((\ref%
{2.96})). Thus, the normalization of the wave function can only be found by
means of a fine tuning on the $\omega _{i} $ 's.

In the case of CBW with $\lambda =\mu $ one can see from (\ref{2.94}) and (%
\ref{2.95}) that the coupling $\omega _{3}~\phi \chi ~$contributes to a
constant for the behavior of $-2\eta \int^{r}F(r^{\prime })dr^{\prime }$ for
$r\rightarrow \pm \infty \,$. In fact, we have found \ that%
\begin{equation}
\exp (-2\eta \int^{r}F(r^{\prime })dr^{\prime })\sim \left\{
\begin{array}{c}
\exp (-2\eta a\omega _{2}r),~~r\rightarrow +\infty \\
\exp (-2\eta a\omega _{1}r),~~r\rightarrow -\infty%
\end{array}%
\right. ~,  \label{36}
\end{equation}%
whilst $e^{-A(r)}\sim e^{4a^{3}\mu |r|/9}$ for $r\rightarrow -\infty $. Then
by choosing $\eta >0$, $\omega _{2}>0$ and $\omega _{1}<-2a^{2}\mu /9\eta $
one has localized left-handed massless fermion. On Figure 4 we show the
profile for the $r$-dependent part of the coefficient of the spinor for $%
\omega _{1}=-1/3,-1,-3$ and $\omega _{3}=0$, $\eta =$ $\omega _{2}=a=\mu =1$%
. As a conclusion, the coupling of fermions with the field $\phi (r)$ is
relevant to provide localized massless fermions, but $\omega _{1}$ should be
close to $-2a^{2}\mu /9\eta $ to insure a sharp localization in the core of
the wall, since the peak of $\alpha _{L_{\ 0}}(r)$ dislocates to the right
of the core of the wall as $|\omega _{1}|$ increases. Moreover, the dominant
contribution for $~r\rightarrow +\infty $ comes from the coupling to the
field $\chi (r)$, which describes the internal structure of the brane, and
we have find that the coupling constant $\omega _{2}$ must be positive in
order to insure the normalization of the wave function associated to the
massless fermion. We have also checked that the coupling $\omega _{3}~\phi
\chi $ contributes to dislocate the peak of $\alpha _{L_{\ 0}}(r)$ from the
core of the wall and that this coupling does not insure the fermion
localization by itself, as has already been shown\ in \cite{benito}. Our
results are in complete agreement with those found in reference \cite{china}.

In the case of CBW with $\lambda =4\mu $ one has the following asymptotic
behaviors%
\begin{eqnarray}
\exp (-2\eta \int^{r}F(r^{\prime })dr^{\prime }) &\sim &\left\{
\begin{array}{c}
\exp (-4\eta a\omega _{2}r),~~r\rightarrow +\infty \\
\exp (-4\eta a\omega _{1}r),~~r\rightarrow -\infty%
\end{array}%
\right. ~,  \nonumber \\
e^{-A(r)} &\sim &\left\{
\begin{array}{c}
\mathrm{const.},~~\ \ \ \ \ \ \ \ \ \ \ \ r\rightarrow +\infty \\
\exp (-\frac{16a^{3}\mu r}{9}),~~r\rightarrow -\infty%
\end{array}%
\right. .  \label{38}
\end{eqnarray}%
such that one can choose $\eta >0$, $\omega _{2}>0$ and $\omega
_{1}<-4a^{2}\mu /9\eta $ in order to have normalizable fermion zero modes
trapped inside the wall.

We have also considered the localization of massless fermions on DBW with $%
\lambda =4\mu $ by taking the solutions (\ref{2.91}) and (\ref{2.92}). We
have found that smooth normalizable solutions are obtained by setting $%
\omega _{2}=\omega _{3}=0$ and $\omega _{1}>8a^{2}\mu /9\eta $. We show in
Figure 5 the behavior of $\alpha _{L_{\ 0}}(r)$ for two different values of $%
c_{0}$, $c_{0}\ll 1/16a^{2}$ and $c_{0}\lessapprox 1/16a^{2}$. One can note
that the behavior of $\alpha _{L_{\ 0}}(r)$ does not follow the behavior of
the brane. Whilst Figure 3 shows the two-kink solutions, which indicates the
brane splitting, Figure 5 shows that the left handed massless fermion is
likely to be seen in the bulk between the two walls as $c_{0}$ approaches
the critical value. Although this seems to be a democracy in the sense that
there is no preferable wall for the fermion to be trapped in, it also brings
a paradox since the fermion lives in the bulk between the two walls
preventing it to be measured. We have been looking for a solution to this
apparent paradox. We have found that a coupling, reminiscent from
supersymmetry, namely%
\begin{equation}
F(\phi ,\chi )=\omega _{1}\frac{W_{\phi \phi }W_{\phi }+W_{\phi \chi
}W_{\chi }}{W_{\phi }}=\omega _{1}\frac{\phi ^{\prime \prime }}{\phi
^{\prime }},
\end{equation}%
with $\omega _{1}<-2a^{2}/9\eta ~$can afford localized massless fermion with
definite chirality whose $r$-dependent behavior follows the brane splitting
by exhibiting a sharp localization in the cores of the walls as can be seen
in Figure 6. In this scenario the massless fermion can be found in both
walls simultaneously and the probability density for the fermion to be found
between both walls diminishes as the walls are far apart from each other.

\section{Conclusions}

In this work we analyze the localization of massless fermions on degenerate
and critical Bloch branes. The calculations are done directly on the
original extra-dimensional spatial variable, that is, the physical
coordinates, in contrast to some works that resort to the transformation of
variables $dz=e^{-A(r)}dr$, that is, the conformal coordinates, in order to
obtain effective time-independent Schr\"{o}dinger equations, for the $z$%
-dependent parts of the spinors, with eigenvalue equals to the fermion mass.
With such a procedure the effective Schr\"{o}dinger equation allows us to
see the emergence of a possible Kaluza-Klein (KK) tower of massive fermions.
On the other hand, the appearance of possible non-localized KK modes is not
evident from the effective Schr\"{o}dinger equations in the physical
coordinate $r$, equations (\ref{23})-(\ref{26}). This is due to the fact
that the mass contributes to the effective potentials, i.e., for each mass
one has a different potential, such that one should look for scattering
states, or even massive\ fermion localized states, by means of an effective
Schr\"{o}dinger equation with eigenvalue zero. Notwithstanding, one does not
have to resort to an effective Schr\"{o}dinger equation in order to analyze
the existence of normalizable localized states of massless fermions.

It is important to remark that the interaction term of fermions with the
scalar field is crucial to the correct localization of fermions on the
brane. The most natural coupling is the Yukawa one,\ namely, $\phi \bar{\Psi}%
\Psi $, used originally in \cite{jackiw-rebbi} to explain the charge
fractionization by a soliton background, and stablished in \cite{rubakov1}
to illustrate localization of fermions in a domain-wall. In fact, the Yukawa
coupling is the simplest one, but such a choice comes naturally if one has
in mind the potential $\frac{\lambda }{2}(\phi ^{2}-1)^{2}$, whose BPS
solution is the soliton background that traps the fermions. In this case,
the Yukawa coupling entails a $N=1$ supersymmetry (SUSY) in the
fermion-boson system, once the superpotential is $W(\phi )=$ $\sqrt{\lambda }%
\phi (\frac{\phi ^{2}}{3}-1)$ and $d^{2}W/d\phi ^{2}=2\sqrt{\lambda }\phi $.
In the context of branes in a warped space-time the issue of SUSY, that is
supergravity, is much more complicate, moreover when it comes with two
scalar fields, which is the case of Bloch branes in a model which supports a
variety of soliton solutions. We have shown that the general coupling $%
(\omega _{1}\phi +\omega _{2}~\chi +\omega _{3}~\phi \chi )\bar{\Psi}\Psi $
guarantees the localization of massless fermions, for a range of the
coupling constants $\omega _{1}$, $\omega _{2}$ and $\omega _{3}$, and that
the cross term $\phi \chi $ does not provide fermion localized states by
itself, in agreement with previously reported calculations \cite{china},
\cite{benito}. The general coupling also works even when one has two-kink
solutions, but the $r$-dependent part of the wave-fuction is peaked just in
the middle of the region between the walls. Such a behavior is not the
desirable one, once it signalizes that the fermions would not be observed
inside the branes. We have proposed another coupling between fermion and
scalar fields which provides the correct localization of fermionic
zero-modes inside the branes. The chosen coupling seems to come as a
reminiscent of a SUSY model. This last issue has been analyzed  in \cite%
{hottetal}.

\bigskip

\textbf{Acknowledgements: }The authors are grateful to CNPq and CAPES for
partial financial support. MBH thanks to Luis Benito Castro for discussion
on questions concerning localization of fermions.

\newpage
\begin{figure}[tbp]
\begin{center}
\begin{minipage}{20\linewidth}
\epsfig{file=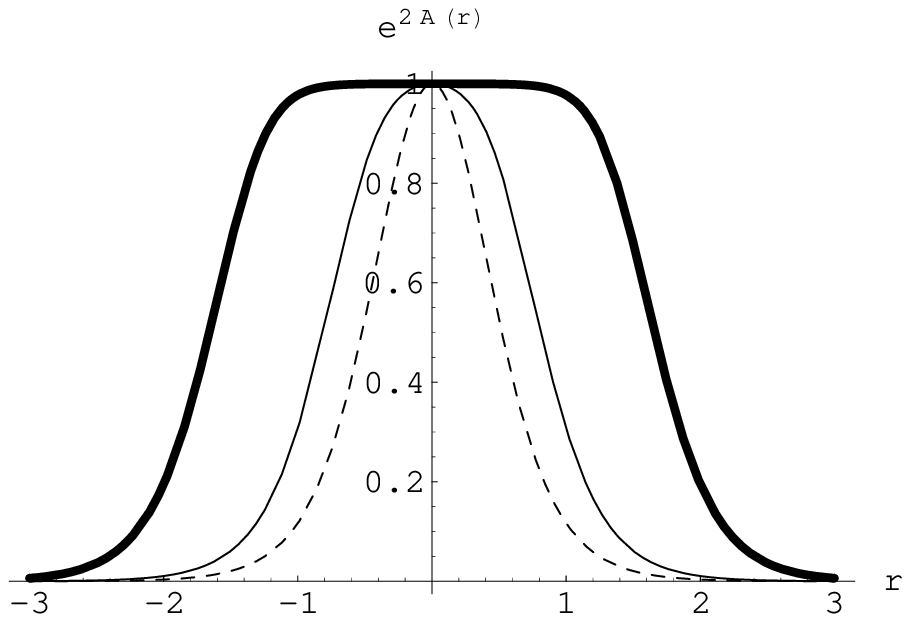}
\end{minipage}
\end{center}
\caption{Warp factor in the case of DBW with $\protect\lambda =4\protect\mu $%
: $c_{0}=0$ (dashed line), $c_{0}=1/17$ (thin solid line), $c_{0}=1/16.001$
(thick solid line) with $a=\protect\mu =1$.}
\label{fig:Fig.1}
\end{figure}

\begin{figure}[tbp]
\begin{center}
\begin{minipage}{20\linewidth}
\epsfig{file=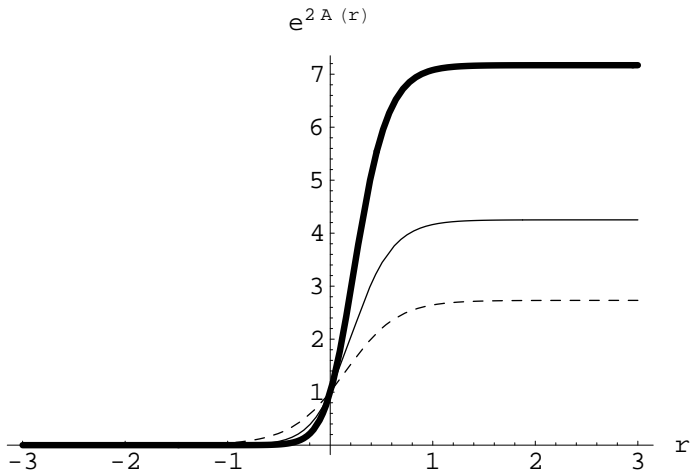}
\end{minipage}
\end{center}
\caption{Warp factor in the case CBW with $\protect\lambda =4\protect\mu $
for $\protect\mu =1$ and $a=1$ (dashed line), $a=1.2$ (thin solid line), $%
a=1.4$ (thick solid line).}
\label{fig:Fig.2}
\end{figure}

\begin{figure}[tbp]
\begin{center}
\begin{minipage}{20\linewidth}
\epsfig{file=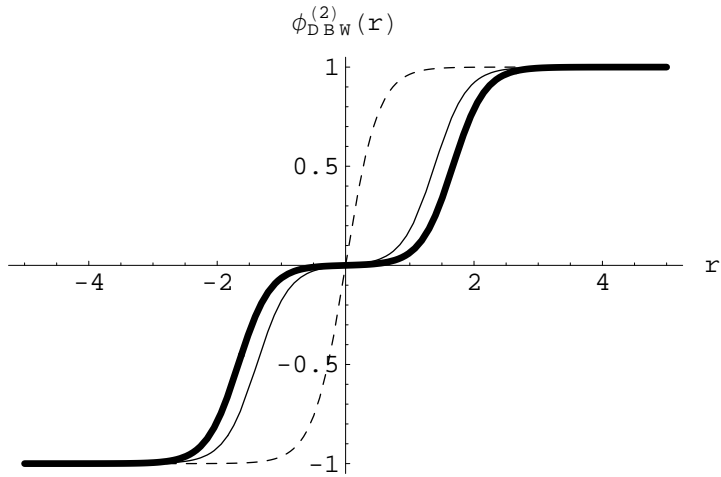}
\end{minipage}
\end{center}
\caption{Two-kink solution (DBW) with $\protect\lambda =4\protect\mu $: $%
c_{0}=0$ (dashed line), $c_{0}=1/16.001$ (thin solid line), $c_{0}=1/16.0001$
(thick solid line) with $a=\protect\mu =1$.}
\label{fig:Fig.3}
\end{figure}

\begin{figure}[tbp]
\begin{center}
\begin{minipage}{20\linewidth}
\epsfig{file=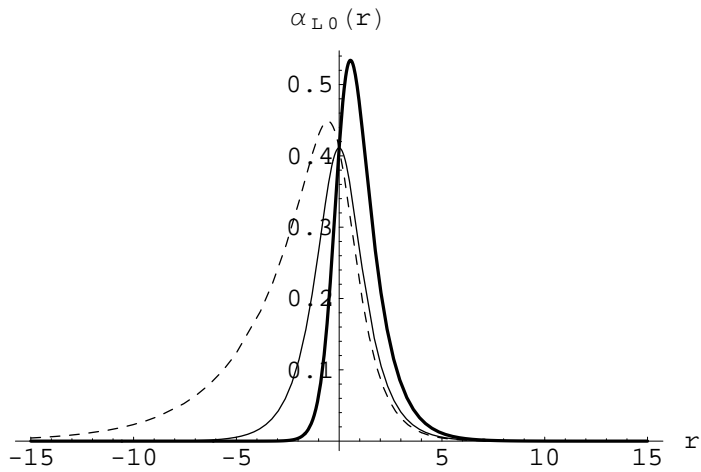}
\end{minipage}
\end{center}
\caption{$\protect\alpha _{L0}(r)$ in the case of CBW with $\protect\lambda %
=4\protect\mu $ and $\protect\omega _{2}=1$, $\protect\omega _{3}=0$, $%
\protect\omega _{1}=-1/3$ (dashed line), $\protect\omega _{1}=-1$ (thin
solid line) and $\protect\omega _{1}=-3$ (thick solid line); $a=\protect\mu %
=1$.}
\label{fig:Fig.4}
\end{figure}

\begin{figure}[tbp]
\begin{center}
\begin{minipage}{20\linewidth}
\epsfig{file=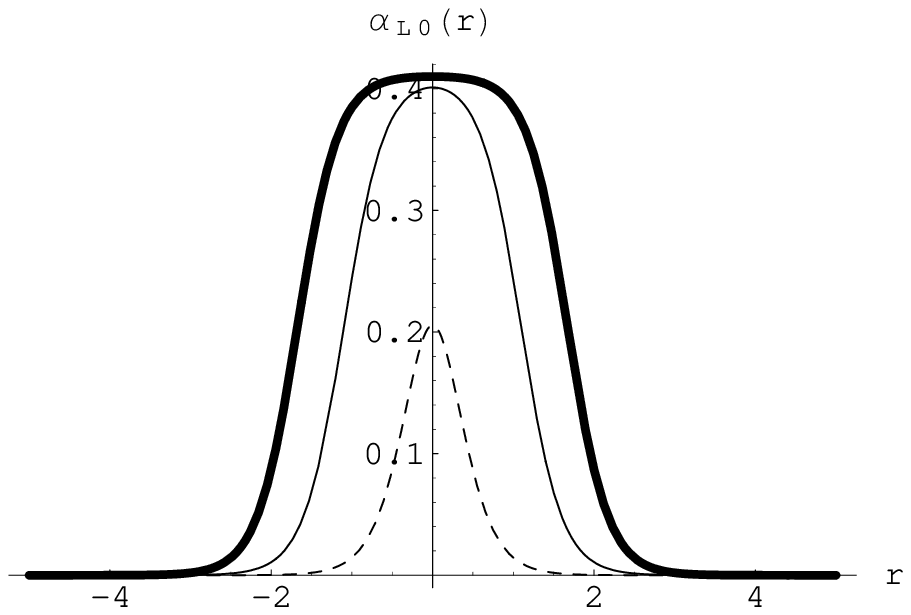}
\end{minipage}
\end{center}
\caption{$\protect\alpha _{L0}(r)$ in the case of DBW ($\protect\lambda =4%
\protect\mu $) and $F=\protect\omega _{1}\protect\phi $ for $\protect\omega %
_{1}=4$, $c_{0}=0$ (dashed line), $c_{0}=1/16.01$ (thin solid line)and $%
c_{0}=1/16.0001$ (thick solid line); $a=\protect\mu =\protect\eta =1$.}
\label{fig:Fig.5}
\end{figure}

\begin{figure}[tbp]
\begin{center}
\begin{minipage}{20\linewidth}
\epsfig{file=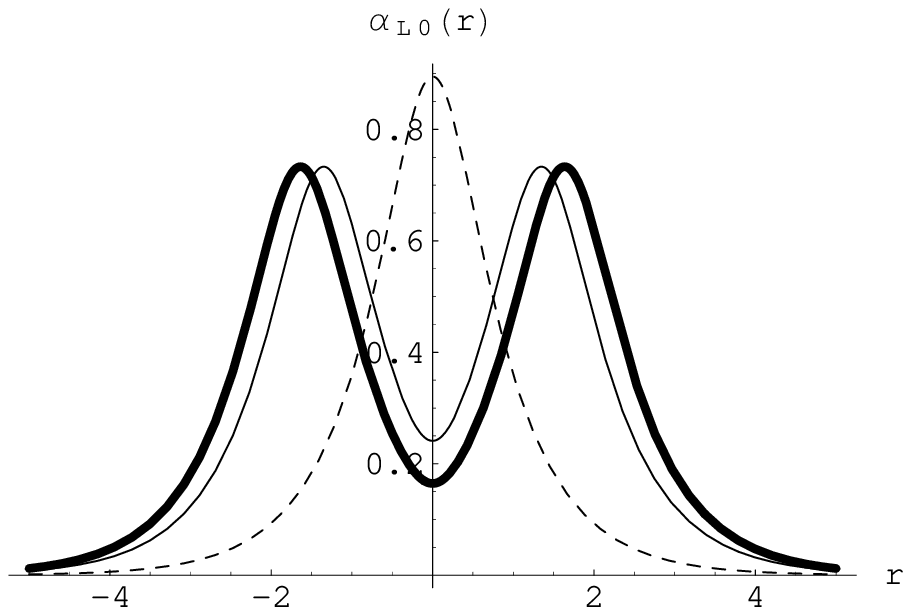}
\end{minipage}
\end{center}
\caption{$\protect\alpha _{L0}(r)$ in the case of DBW ($\protect\lambda =4%
\protect\mu $) and $F=\protect\omega _{1}\protect\phi ^{\prime \prime }/%
\protect\phi ^{\prime }$ for $\protect\omega _{1}=-1/3$, $c_{0}=0$ (dashed
line), $c_{0}=1/16.001$ (thin solid line)and $c_{0}=1/16.0001$ (thick solid
line); $a=\protect\mu =\protect\eta =1$.}
\label{fig:Fig.6}
\end{figure}

\end{document}